\documentclass[11pt]{article}

\usepackage[preprint]{acl}

\usepackage{times}
\usepackage{latexsym}
\usepackage[T1]{fontenc}
\usepackage[utf8]{inputenc}
\usepackage{microtype}
\usepackage{graphicx}
\usepackage{booktabs}
\usepackage{array}
\usepackage{multirow}

\title{Language-Specific Gaps in AI Safety Training Datasets}

\author{
  Chialuka Prisca-Mary Onuoha, Bright Etornam Sunu, Rashidat Sikiru \\
  Black in AI Safety \& Ethics (BASE) \\
  \texttt{chialukaonuoha@gmail.com}, \texttt{sunumacbright@gmail.com}, \texttt{rasheedatsikiru@gmail.com}
}

\begin{document}
\maketitle

\begin{abstract}
Large language model providers routinely cite multilingual safety benchmarks spanning a dozen or more languages as evidence that their models are safe for non-English-speaking users. We show that these collection-level coverage claims frequently do not survive inspection at the level of an individual language. Auditing 21 resources across 25 language slices, of which 20 count as datasets under our counting rules, spanning three languages chosen to represent low- (Hausa), mid- (Swahili), and high-resource (French) tiers, we find that gaps in provenance, annotation reliability, access, harm-taxonomy coverage, and data reuse recur in patterns that partially, but not fully, track resource level. Using a controlled within-pipeline comparison, we show a Hausa-language slice falling below its own paper's translation-quality acceptance threshold while the same pipeline's Swahili output clears the same bar comfortably; this is evidence that these gaps are measurable and addressable, not inherent. We further show that self-harm and sexual-content categories have no native-language coverage in either African-language tier we studied, a total rather than gradated gap that a purely resource-level account does not predict. We connect these findings to a documented, persistent asymmetry in multilingual jailbreak robustness (single-turn attacks largely mitigated, multi-turn attacks still effective), arguing that this asymmetry is structurally consistent with where our audit finds training and evaluation data thinnest. We contribute a reusable slice-level audit methodology, a cross-tier empirical comparison, and concrete recommendations for dataset creators, model providers, and venues aiming to make ``multilingual coverage'' claims verifiable rather than merely stated.\footnote{Dataset: \url{https://huggingface.co/datasets/ChialukaOnuoha/safety-slice-audit}}
\end{abstract}

\section{Introduction}
\label{sec:intro}

Large language models are increasingly deployed to users who speak languages other than English, and model providers routinely cite multilingual safety evaluations as evidence that their models are safe for these users. A model card or technical report will state that a model has been ``evaluated across N languages including [low-resource language],'' and a reader reasonably infers that this evaluation approaches the rigor of the model's English-language safety evaluation. These claims typically rest on a small number of widely-cited resources, datasets such as Aya Red-teaming \citep{aakanksha2024multilingual}, AfriHate \citep{muhammad2025afrihate}, AfriSenti \citep{muhammad2023afrisenti}, and MultiJail \citep{deng2024multilingual}, each of which reports coverage across a dozen or more languages at the collection level.

We show that this inference is frequently unwarranted. Working from a structured audit of 21 resources across 25 language slices spanning three languages selected to represent low- (Hausa), mid- (Swahili), and high-resource (French) tiers, we find that the aggregate coverage claims made by multilingual datasets routinely dissolve on inspection of the individual language slice. We apply a set of explicit counting rules developed for this study (\S\ref{sec:counting}): a derivative re-evaluation of an existing prompt set does not count as an independent dataset, and an unreleased, proprietary resource is not counted as an available one. Under those rules, this candidate pool resolves to 20 datasets across 25 audited language slices. A dataset that reports ``14 languages, 20,000 examples'' may, for a specific low-resource language, resolve to a few hundred machine-translated prompts validated by a single annotator, evaluated with no measure of inter-annotator agreement, and distributed under a license or access restriction that a downstream developer is unlikely to notice before citing the dataset as evidence of coverage.

Critically, we find that these gaps are not evenly or randomly distributed. They cluster by resource level, and they cluster in specific, recurring forms: provenance is systematically overstated, with content described as ``native-authored'' that our audit traces to English-authored, machine-translated material; inter-annotator agreement is disproportionately unreported for the lowest-resource language in a collection, often for a defensible ethical reason (such as protecting annotators from prolonged exposure to abusive content) that nonetheless leaves the lowest-resource slice least verified; access is more frequently gated or request-only; and the same underlying raw text is repeatedly recycled across nominally distinct ``datasets'' through chains of re-annotation, inflating apparent corpus diversity without adding independent evidence. A single 2026 pipeline that processed both Hausa and Swahili through identical translation and validation steps offers a clean, quantified illustration: the Hausa policy-material slice scored below the paper's own translation-quality acceptance threshold, while the Swahili slice comfortably cleared the same bar, evidence that the gap between these two languages is measurable, and, by construction, not attributable to differences in how carefully either language was handled at the level of process design.

The practical stakes of this pattern are not abstract. \citet{deng2024multilingual} found approximately an 83\% unsafe-response rate for Swahili jailbreak attempts against a 2023-era model; a 2026 replication \citep{marx2026multilingual} found that this specific attack surface (single-turn, translation-based jailbreaks) had been largely closed, but that multi-turn conversational jailbreaks in the same low-resource languages remained highly effective, with harmful-response rates ranging from roughly 42\% to 71\% depending on model and language. Providers appear to have patched the vulnerability that was easy to benchmark and left largely open the vulnerability that requires the kind of sustained, native-reviewed, non-machine-translated evaluation data that our audit finds is disproportionately absent for lower-resource languages. The datasets that would be needed to catch the remaining gap are, by our audit, the ones least likely to exist.

This paper makes the following contributions:
\begin{enumerate}
    \item A reusable audit methodology for assessing individual language slices of multilingual AI safety datasets along two axes: data quality (provenance, annotation, agreement) and artifact accessibility (license, access, documentation); rather than treating a dataset's headline collection-level statistics as representative of any one language within it.
    \item A structured comparative audit of 20 datasets across three resource tiers (Hausa/low, Swahili/mid, French/high), yielding a typology of recurring gap types and evidence that gap type and severity correlate with resource level for most, though not all, of the dimensions we measure.
    \item A quantified, controlled case study demonstrating a resource-level quality gap within a single evaluation pipeline, isolating translation and annotation quality from confounds of dataset origin or methodology.
    \item A discussion connecting dataset-level gaps to downstream safety outcomes, using the persistence of multi-turn jailbreak vulnerability in low-resource languages as evidence that dataset gaps are not merely a documentation problem but a live safety problem.
    \item A set of recommendations for dataset creators, model providers, and venues, aimed at making ``multilingual coverage'' claims verifiable at the language-slice level rather than the collection level.
\end{enumerate}

The remainder of the paper proceeds as follows. \S\ref{sec:related} situates this work relative to prior work on dataset documentation, resource-level taxonomies, and multilingual safety evaluation. \S\ref{sec:scope} defines the scope of ``AI safety'' used in this study and justifies our choice of Hausa, Swahili, and French as tier anchors. \S\ref{sec:methodology} details our audit methodology. \S\ref{sec:corpus} describes the corpus of datasets reviewed. \S\ref{sec:findings} presents our findings, organized by gap type. \S\ref{sec:discussion} discusses implications for model developers and benchmark consumers. \S\ref{sec:recommendations} offers recommendations, and \S\ref{sec:limitations}/\S\ref{sec:ethics} cover limitations and ethical considerations.

\section{Related Work}
\label{sec:related}

\paragraph{Dataset documentation and transparency.} Our work builds directly on the tradition of dataset documentation initiated by Datasheets for Datasets \citep{gebru2021datasheets} and Data Statements for NLP \citep{bender2018data}, both of which argue that a dataset's fitness for a given use cannot be assessed from its headline description alone and must instead be evaluated against a structured set of questions about provenance, collection, annotation, and intended use. Several of the datasets in our corpus explicitly adopt the data-statement format (e.g., XTREMESPEECH; \citealp{maronikolakis2022listening}, includes a full data statement as an appendix). Our contribution relative to this line of work is not a new documentation template, but an audit methodology: we treat the data statement or dataset card as a set of claims to be checked against the paper's own tables, the linked repository's actual contents, and where a collection spans multiple languages against the specific slice relevant to a given language, rather than accepting collection-level documentation as sufficient for language-level claims.

\paragraph{Resource-level taxonomies.} We adopt the broad conceptual framing of \citet{joshi2020state}, who classify the world's languages into resource tiers based on the quantity of labeled and unlabeled data available and document the severe skew of NLP research effort toward a small number of high-resource languages. Several of the datasets in our corpus operationalize resource level using a specific quantitative proxy: the language's share of the CommonCrawl corpus, following \citet{lai2023chatgpt} as adopted in \citet{deng2024multilingual} and again in \citet{marx2026multilingual}, which classify a language as low-resource below a 0.1\% CommonCrawl share. We note in \S\ref{sec:scope} that this quantitative convention is not universally applied across our corpus, and that datasets disagree in practice about whether Swahili belongs in a ``low'' or ``mid'' tier: a disagreement that is itself informative and that we surface explicitly rather than resolve by fiat.

\paragraph{Multilingual and low-resource hate speech / offensive-content resources.} A substantial recent literature has produced dedicated hate-speech and offensive-content datasets for African languages specifically, motivated by the observation that general-purpose multilingual safety data underrepresents the continent's languages. This includes AfriHate \citep{muhammad2025afrihate} and AfriSenti \citep{muhammad2023afrisenti}, both products of the Masakhane/HausaNLP research community, as well as language-specific efforts such as the Hausa Offensive Content dataset \citep{adam2023detection} and community-embedded collection frameworks such as XTREMESPEECH \citep{maronikolakis2022listening}, which recruits local fact-checkers rather than external annotators specifically to address the demographic mismatch between annotator pools and affected communities documented by \citet{sap2019risk} and \citet{founta2018large}. Our audit draws heavily on this literature as source material but asks a different question of it: not ``does a good-faith, native-reviewed Hausa or Swahili hate-speech dataset exist'' (it generally does) but ``how does its evidentiary strength compare, slice for slice, to the high-resource language(s) it is typically evaluated alongside.''

\paragraph{Multilingual jailbreak and safety-training-gap literature.} A parallel body of work has established, empirically, that safety alignment transfers unevenly across languages. \citet{deng2024multilingual} introduced MultiJail and showed low-resource-language jailbreak susceptibility roughly three times that of high-resource languages under a translation-based attack; \citet{yong2024lowresource} demonstrated a similar effect specifically for low-resource languages against GPT-4; \citet{wang2024all} extended this analysis across a broader language set under the heading ``All Languages Matter.'' \citet{marx2026multilingual} revisit this question against more recent (2026-era) commercial models and find that single-turn translation attacks have been substantially mitigated, while multi-turn conversational attacks remain effective, with translation quality itself measured via BERTScore, METEOR, and BLEU against back-translation, the dominant factor explaining residual cross-language variance in attack success. This last finding is directly load-bearing for our argument: if translation/annotation quality is the mechanism driving safety-training-gap severity, then an audit of the quality of the training and evaluation data available per language (which is what we conduct) is a more direct diagnostic than language identity alone.

\paragraph{Government and multi-institution safety evaluation efforts.} Most recently, national and multi-national AI Safety Institutes have begun publishing joint testing exercises that include non-English, non-hate-speech safety domains, for instance, the Third International Joint Testing Exercise on agentic safety (fraud and sensitive-information leakage), which includes a Kenya AISI-contributed Kiswahili component \citep{intlnetwork2025agentic}. We discuss this resource in \S\ref{sec:scope} primarily as a boundary case that clarifies the scope of ``AI safety'' adopted in this paper, since its harm taxonomy (fraud, privacy leakage) falls outside the hate-speech/harassment/self-harm/extremism/sexual-content/misinformation framework used elsewhere in our corpus.

\paragraph{Gap in the literature.} To our knowledge, no prior work systematically audits language-slice-level evidentiary quality across a set of multilingual AI safety datasets using a shared, resource-tier-stratified methodology, or connects the resulting typology of gaps directly to a documented, persistent downstream safety outcome (residual multi-turn jailbreak vulnerability). This is the gap this paper addresses.

\section{Scope and Definitions}
\label{sec:scope}

\subsection{Defining ``AI safety'' for this study}

``AI safety'' is used inconsistently across the literature we review, spanning at least three distinct senses: (a) content safety: detection or evaluation of hateful, harassing, self-harm-related, sexually explicit, extremist, or misinformation content; (b) adversarial/jailbreak safety: evaluation of whether a model can be induced to produce content it is trained to refuse; and (c) agentic/operational safety: evaluation of whether a model, acting with tool access, can be induced into fraud, privacy leakage, or other operationally harmful behavior.

This study centers on senses (a) and (b): datasets and benchmarks addressing hate speech, harassment, self-harm, extremism, sexual content, and misinformation, together with jailbreak/red-teaming resources that probe a model's willingness to produce content in these categories. We adopt this six-category harm framework, adapted from the Trust and Safety Professional Association (TSPA) harm taxonomy, as our primary lens because it is the framework against which the majority of our corpus can be meaningfully mapped, even where a given dataset's own native taxonomy uses different labels (e.g., XTREMESPEECH's derogatory/exclusionary/dangerous scale, or AfriHate's hate/abusive/neutral scale).

We explicitly exclude sense (c), agentic/operational safety, from primary analysis. We encountered one dataset in this domain during our review, the Third International Joint Testing Exercise's Kiswahili deep-dive (agentic fraud and sensitive-information-leakage testing), and rather than silently omit it, we use it in \S\ref{sec:boundary} as a boundary case: it is native-reviewed, partially native-authored, and methodologically serious, but its harm taxonomy (fraud, PII leakage) does not overlap with any of our six tracked categories. We treat this as informative rather than merely inconvenient: it illustrates that ``safety'' resources for a given low-resource language may exist and be well-constructed while still leaving the content-safety and jailbreak gap this paper is concerned with completely unaddressed. A downstream reader who encounters this resource and concludes ``Kiswahili safety has been evaluated'' would be missing this distinction.

\subsection{Resource tiers and the choice of Hausa, Swahili, and French}

We stratify our corpus into three resource tiers: low (Hausa), mid (Swahili), and high (French); chosen to allow direct, controlled comparison of gap severity across the resource spectrum while remaining tractable within a single audit.

We do not treat this stratification as a settled, independently-derived classification; rather, we surface it as itself a point of disagreement in the literature and are transparent about that disagreement. Several datasets in our corpus that explicitly operationalize resource level via CommonCrawl-share thresholds \citep{deng2024multilingual,marx2026multilingual} classify Swahili as low-resource (CommonCrawl share below the 0.1\% threshold used in both papers), placing it in the same tier as Hausa by that specific metric. Other resources in our corpus and our own working categorization prior to this paper treat Swahili as a mid-resource language, reflecting its status as one of the most widely spoken languages in Africa (Swahili, Amharic, and Hausa are noted as the three largest African languages by speaker count in \citealp{muhammad2023afrisenti}) and the comparatively larger volume of curated NLP resources available for it (e.g., it appears in AfriHate, RTP-LX, XTREMESPEECH and UbuntuGuard, a wider set of benchmarks than Hausa receives). We retain the low/mid/high framing in this paper because it reflects the practical resource picture (number and diversity of available datasets, not raw web-crawl token share) that a developer would actually encounter when trying to source safety training data for a given language, but we flag this tension explicitly here rather than let it operate as an unexamined assumption, and we return to it in \S\ref{sec:findings} where we show that Swahili's outcomes in our audit are in fact intermediate between Hausa's and French's on most of our gap dimensions, which is some independent evidence in favor of the mid-tier framing on the dimensions this paper actually measures.

Hausa, Swahili, and French were further selected because each has a non-trivial presence across multiple datasets in our corpus, permitting within-language cross-dataset comparison (e.g., Hausa across AfriHate, AfriSenti, HOC, TukaBench, UbuntuGuard, and the LSR benchmark) as well as the resource-tier cross-language comparison that is the paper's central contribution. French additionally serves a specific comparative function: several datasets in our corpus (Aya Red-teaming, RTP-LX, XSafety, CultureGuard, Jigsaw) were constructed with French positioned as a genuinely first-class, natively-supported language rather than an afterthought extension, giving us a credible ``what good coverage looks like'' reference point rather than an arbitrary high-resource baseline.

\subsection{Defining a ``gap''}
\label{sec:gapdef}

We use ``gap'' throughout this paper to refer to any point of divergence between a dataset's collection-level coverage claim and the evidentiary quality actually available for a specific language slice within it. Our audit methodology (\S\ref{sec:methodology}) operationalizes this across the following gap categories, which recur across our corpus and structure our findings in \S\ref{sec:findings}:

\begin{itemize}
    \item \textbf{Provenance gaps}: content described as native-authored, transcreated, or otherwise culturally grounded that traces, on inspection of the paper's own methodology section, to machine translation from English with limited or no native post-editing.
    \item \textbf{Annotation/agreement gaps}: absence of reported inter-annotator agreement, or agreement reported only at the collection level rather than the language-slice level, particularly where this absence is attributable to resource constraints rather than a considered methodological choice.
    \item \textbf{Access and licensing gaps}: data that is gated, request-only, or unlicensed, such that a claim of ``publicly available'' coverage does not translate into practical reproducibility.
    \item \textbf{Taxonomy-coverage gaps}: harm categories absent from a given language's evaluation relative to the categories nominally covered by the collection as a whole, or relative to our six-category framework (\S\ref{sec:scope}).
    \item \textbf{Reuse and double-counting gaps}: apparent corpus diversity that resolves, on provenance tracing, to repeated re-annotation of the same underlying raw text across nominally distinct datasets.
    \item \textbf{Quantified quality gaps}: cases, rarer but especially valuable, where a single study's own reported metrics (translation-quality scores, judge-human discrepancy rates) allow a direct, controlled quality comparison across languages within one pipeline.
\end{itemize}

Each dataset in our corpus is assessed against all six categories; \S\ref{sec:findings} presents the resulting pattern of findings organized by category, with resource tier as the cross-cutting comparison within each.

\section{Audit Methodology}
\label{sec:methodology}

This section sets out the instrument and procedure used to assess each dataset. Our methodological claim is narrow but consequential: the appropriate unit of analysis for a multilingual safety dataset is not the dataset but the language slice, and once that unit is adopted, a set of gaps becomes visible that collection-level documentation systematically conceals.

\subsection{Unit of analysis: the language slice}

A multilingual dataset is not a single artifact of uniform quality. It is a collection of per-language subsets that may differ in how they were produced, how many items they contain, who reviewed them, and whether agreement was ever measured. A collection whose headline describes professional transcreation by native speakers may nonetheless contain individual language slices that were machine-translated in a later release and validated by one person; a collection reporting inter-annotator agreement may report it only in aggregate, leaving the lowest-resource slice unverified.

We therefore treat the language slice as the unit of audit. Each row of our instrument records one language within one dataset. A dataset covering both Hausa and Swahili yields two rows, each assessed independently, and the two rows may legitimately reach different conclusions. This is the operational meaning of the distinction our findings depend on, and it is also the discipline that exposed several errors in our own early annotations, discussed in \S\ref{sec:verification}.

\paragraph{Consequence for reporting.} Because one dataset can produce several rows, the number of rows in our audit is not the number of datasets we reviewed. We state both figures separately and define the counting rule in \S\ref{sec:counting}, so that no claim about corpus size is inflated by slice multiplicity.

\subsection{The audit instrument}

The instrument is a thirty-one-field schema applied identically to every slice. It was developed iteratively: an initial draft derived from the question set in Datasheets for Datasets \citep{gebru2021datasheets} was extended with fields that our first annotation pass showed to be load-bearing, and further extended with a group of provenance-depth and usability fields after early review. Fields are grouped into seven families, summarized in Table~\ref{tab:instrument}.

\begin{table*}[t]
  \centering\small
  \begin{tabular}{p{2.7cm}p{6.5cm}p{5.8cm}}
    \toprule
    \textbf{Field group} & \textbf{Fields} & \textbf{What it establishes} \\
    \midrule
    Identification & Dataset name; year/version; source paper (DOI/arXiv); dataset link (repository); authors; model(s) evaluated & That the artifact is uniquely identifiable and that the paper and the data are separately locatable \\
    Language coverage & Language(s) covered; resource level; monolingual vs.\ multilingual & Whether the target language is a first-class subject or one slice of a larger collection \\
    Provenance & Method of creation (native / translated / machine-translated / transcreated / synthetic / mixed); translation direction; provenance evidence & How this language's items were actually produced, evidenced by the paper's own methodology section \\
    Harm coverage & Harm categories covered; harm category gaps; native taxonomy (verbatim); global vs.\ local harm & Coverage against our six-category framework, and the authors' own conceptualisation before mapping \\
    Annotation \& culture & Who annotated it; native review of this language; annotator count and background; IAA for this slice; dialect and cultural handling; script; code-switching coverage & Evidentiary strength of the labels, and whether cultural and orthographic variation is modelled \\
    Artifact characteristics & Format/structure (prompt-only, multi-turn, prompt+response, blocklist); size for this language; domain/source; entries unit & Fitness for a given downstream use, and slice magnitude rather than collection magnitude \\
    Usability & License/usage rights; access status; annotation guidelines public; reported per-language performance & Whether a coverage claim converts into practical, reproducible availability \\
    Audit record & Rationale for the call; quality concerns; include / exclude / flag & The judgment trail, so every decision is attributable and reversible \\
    \bottomrule
  \end{tabular}
  \caption{The audit instrument, grouped by field family. Each of the thirty-one fields is applied to every language slice, and each field carries a one-sentence rationale in the instrument itself so that the schema is self-documenting.}
  \label{tab:instrument}
\end{table*}

Three design choices in this schema are worth making explicit, because they encode the paper's argument. First, provenance and size are both scoped to the slice rather than the collection. Second, native review of the specific language is recorded as a separate structured field rather than folded into a general annotation field, because authorship and review diverge in both directions: a translated slice may receive genuine native post-editing (TukaBench), while a natively written one may carry no independent native review at all (LSR). Third, the dataset's own harm taxonomy is recorded verbatim, before any mapping to our six categories, so that the authors' conceptualisation is preserved and our mapping remains inspectable.

\subsection{Slice-level verification protocol}
\label{sec:verification}

Every field that could be contradicted by a primary source was checked against three sources in order: the paper's own methodology section and per-language tables; the linked repository or dataset card; and, where a repository was live, the actual contents of the release. Where these disagreed, the paper's per-language table took precedence for claims about how data was made, and the live release took precedence for claims about what is available.

This protocol was not a formality. It changed the record in four distinct ways, each of which generalises into a finding reported in \S\ref{sec:findings}:

\begin{enumerate}
    \item \textbf{Language-assignment correction.} One dataset was initially recorded against Hausa on the strength of a plausible secondary description; the repository's own language list shows that the language is not covered by that dataset, while a second language had in fact been added in a later release and was initially recorded as absent. Both errors were caught only by reading the repository's version history rather than the paper abstract.
    \item \textbf{Provenance correction.} Two datasets whose secondary descriptions imply native authorship state in their own methodology sections that items were written in English and then machine-translated, with post-editing in one case and none reported in the other.
    \item \textbf{Release-state correction.} One dataset's live release contains substantially fewer items for our target language than the paper describes, because components documented in the paper have not been uploaded. Another was announced for release and never shipped.
    \item \textbf{Attribution correction.} One dataset's widely-quoted expert-authorship claim applies to the English seed material only; the paper states explicitly that generating that material is not its contribution.
\end{enumerate}

We record these as methodology rather than as errata because they indicate that the audit procedure is doing work that reading abstracts does not do. A reviewer should be able to see that our findings could not have been obtained from collection-level descriptions.

\subsection{Harm-taxonomy normalisation}
\label{sec:harmtax}

Datasets in our corpus label harm using at least ten mutually non-equivalent schemes, including three-class hate scales, binary offensiveness, eight-dimension toxicity vectors, ten-behaviour jailbreak taxonomies, twelve-category hazard hierarchies, and theme-by-domain policy matrices. Comparing coverage across languages requires a single reference framework, and constructing one requires judgment that the original authors did not necessarily intend. We therefore adopt the six-category framework introduced in \S\ref{sec:scope}, adapted from the Trust \& Safety Professional Association (TSPA) harm taxonomy, state a working definition for each category, and record every mapping decision. Table~\ref{tab:taxonomy} gives the working definitions and representative native labels mapped to each category.

\begin{table*}[t]
  \centering\small
  \begin{tabular}{p{2.1cm}p{6.4cm}p{6.4cm}}
    \toprule
    \textbf{Harm category} & \textbf{Working definition adopted for this study} & \textbf{Representative native labels mapped in} \\
    \midrule
    Hate speech & Content attacking or demeaning a person or group on the basis of a protected or group identity (ethnicity, religion, nationality, gender, disability, political-communal identity). & AfriHate ``hate'' + target attributes; XTREMESPEECH ``exclusionary''; Onyango hate-target labels; Jigsaw ``identity hate''; JBB ``harassment/discrimination'' \\
    Harassment & Targeted abusive, insulting, threatening or degrading conduct directed at an individual, without necessarily invoking group identity. & AfriHate ``abusive''; HOC ``offensive'' (binary); Jigsaw ``insult''/``threat''/``obscene''; RTP-LX ``insult'', ``microaggression''; XTREMESPEECH ``derogatory'' \\
    Self-harm & Content that encourages, instructs, or normalises self-injury, suicide, or disordered behaviour. & RTP-LX ``self-harm''; Aegis 2.0 self-harm hazards (CultureGuard); ALERT self-harm micro-categories; XSafety ``mental health'' (partial) \\
    Extremism & Content promoting violent ideological action, violent organisations, or incitement to communal or political violence. & XTREMESPEECH ``dangerous speech'' (incitement to communal or political violence); ALERT/Aegis extremism hazards \\
    Sexual content & Explicit sexual material, including sexualised depictions of minors and non-consensual sexual content. & RTP-LX ``sexual content''; Aya ``non-consensual sexual content''; JBB ``sexual/adult''; Aegis sexual hazards \\
    Misinformation & Verifiably false or fabricated claims presented as fact, including political disinformation and health misinformation. & PolitiKweli ``fake'' vs.\ ``fact''; UbuntuGuard ``misinformation and disinformation'' theme; JBB ``disinformation''; ALERT misinformation categories \\
    \bottomrule
  \end{tabular}
  \caption{The six-category harm framework, the working definition adopted for this study, and representative native labels mapped into each category. Mappings are recorded per slice in the audit instrument alongside the verbatim native taxonomy.}
  \label{tab:taxonomy}
\end{table*}

Mapping is a documented judgment, not a measurement. Three principles govern it. First, we map only where a native label's own definition, as stated in its source paper, falls within our working definition; we do not infer coverage from a label's name. Second, where a native label spans two of our categories, we record it against both and flag the split, rather than choosing one; the binary offensiveness label in one Hausa dataset, for example, conflates hate speech and harassment and is recorded as such. Third, where a native label falls entirely outside our six categories, we record it as out-of-framework instead of forcing it in, which is what produces the boundary case discussed in \S\ref{sec:boundary}. Because these are judgments, the verbatim native taxonomy is retained in every row so that a reader who disagrees with a mapping can recover the original and re-derive their own.

\subsection{Inclusion and exclusion criteria}

A resource was included if it satisfied both of the following: it contains or evaluates at least one of Hausa, Swahili, or French; and its purpose is content safety or adversarial safety as defined in \S\ref{sec:scope}, meaning toxicity, hate or abusive speech, red-teaming, jailbreak, or content moderation.

The following were excluded, with the reason recorded per row:
\begin{itemize}
    \item General-purpose resources without a safety purpose, including named-entity, part-of-speech, machine-translation, and speech corpora.
    \item Method and model papers that evaluate on an existing benchmark without releasing new data; these are cited as evidence about datasets, not counted as datasets.
    \item Lexicons and wordlists containing no sentences and no labels, which cannot support the provenance or annotation analysis this audit performs.
\end{itemize}

Two categories were included with an explicit flag instead of being excluded. Repurposed resources, where a dataset built for another task is used as safety material, are included and flagged, because their reuse is common practice in this literature and because the reuse itself is a finding: one sentiment corpus in our set deleted offensive content during collection, which compromises exactly the reuse it is subject to. Boundary-domain resources, whose safety purpose is real but whose harm taxonomy falls outside our six categories, are included and analysed separately in \S\ref{sec:boundary} rather than dropped, for the reason given in \S\ref{sec:scope}.

\subsection{Counting rules}
\label{sec:counting}

Because the unit of audit is the slice and the unit of interest for corpus-size claims is the dataset, we adopt three explicit rules.

\begin{enumerate}
    \item \textbf{Datasets are counted once.} A dataset appearing under two languages contributes one dataset and two slices. Per-language results are reported as slices of one dataset, never as distinct datasets.
    \item \textbf{Derivative re-tests are not datasets.} A study that re-translates and re-evaluates an existing prompt set, without releasing a new corpus, is counted as evidence about that dataset, not as an additional dataset, and is cited as such.
    \item \textbf{Re-annotation layers are counted once, at the layer.} Where a dataset is constructed by merging and re-labelling existing corpora, it is counted as one dataset, and its provenance chain is recorded so that the underlying items are not counted again through their original sources.
\end{enumerate}

These rules matter quantitatively. Under rule three alone, one Swahili resource that presents 101,014 labelled items is built from three corpora already in our audit; counting its items as new coverage would roughly quadruple our apparent Swahili corpus. We report both totals, unique datasets and slices audited, wherever a count appears.

\subsection{Assigning gaps}

Each slice was assessed against all six gap categories from \S\ref{sec:gapdef} and assigned one of four levels per category: none, mild, moderate, or severe. Levels are anchored to observable criteria, not impression. For provenance, severity rises as the slice moves from native authorship through transcreation and human translation to unreviewed machine translation or synthetic generation. For annotation, severity rises where slice-level agreement is unreported, where agreement is reported only in aggregate, and where reported agreement falls below conventional thresholds. For access, severity rises through open, gated, request-only, and unreleased. For taxonomy coverage, severity is a count of our six categories with no native coverage. For reuse, severity rises with the number of audited datasets sharing underlying items. For quantified quality, a gap is recorded only where a single study reports a comparable metric across two or more of our languages, which is what makes \S\ref{sec:quantquality} possible.

\paragraph{Reliability of the assignment.} Assignments were made by a single annotator per slice, which is a limitation we state in \S\ref{sec:limitations}. To mitigate it, every assignment carries a written rationale citing the specific paper section, table, or repository state on which it rests, and the flagged rows were reviewed jointly by the author team. We report the anchoring criteria above precisely so that the assignment can be reproduced or contested.

\section{Corpus Under Study}
\label{sec:corpus}

\subsection{Composition and counts}

Our audit comprises 25 language slices drawn from 21 uniquely named resources; a 26th slice, Ubisoft's ToxBuster game-chat work, was reviewed and excluded because its data is not public (\S\ref{sec:boundarydecisions}). Applying the counting rules in \S\ref{sec:counting}, one of the 21 resources is a derivative re-test rather than a dataset, yielding 20 datasets, of which 19 fall within the content-safety and adversarial-safety scope defined in \S\ref{sec:scope} and one is the boundary case analysed in \S\ref{sec:boundary}. Slices are distributed as seven for Hausa, ten for Swahili, and eight for French.

The asymmetry in that distribution is itself the first observation of the audit, and it runs in a direction worth stating plainly: our mid-resource language is represented by more slices than our high-resource language. This does not indicate that Swahili is better served. As \S\ref{sec:findings} shows, Swahili's slice count is inflated by a chain of re-annotations of a shared pool of tweets and by studies that re-evaluate the same prompt set, whereas the French slices are largely independent resources. Slice count is a poor proxy for coverage, which is why the remainder of this section characterises the corpus by provenance and format, not by volume.

Table~\ref{tab:corpus} lists the full audited corpus by tier, with provenance and size recorded at the language-slice level.

\begin{table*}[t]
  \centering\scriptsize
  \begin{tabular}{p{3.3cm}p{0.5cm}p{2.3cm}p{2.1cm}p{5.6cm}}
    \toprule
    \textbf{Dataset (slice)} & \textbf{Tier} & \textbf{Provenance of slice} & \textbf{Format} & \textbf{Slice size / key figure} \\
    \midrule
    \multicolumn{5}{l}{\textit{HAUSA (low)}} \\
    AfriHate \citep{muhammad2025afrihate} & HA & Native-authored & Labelled posts & 6,644 tweets; $\kappa=0.75$ \\
    AfriSenti \citep{muhammad2023afrisenti} & HA & Native-authored & Labelled tweets & 22,155 tweets (14,173 / 2,678 / 5,304); $\kappa=0.66$ \\
    HOC \citep{adam2023detection} & HA & Native-authored & Labelled posts & Size not reported; no IAA \\
    NaijaOffens \citep{aliyu2024beyond} & HA & Native-authored & Labelled tweets & Unreleased; size not reported \\
    TukaBench \citep{akinode2026tukabench} & HA & MT + native post-edit & Prompt-only & 986 described / 300 released; judge--human agreement 69.4\% \\
    UbuntuGuard \citep{abdullahi2026ubuntuguard} & HA & Synthetic + MT & Multi-turn & 1,656 train / 278 test; policy MT 66.37, transcript 93.31 \\
    LSR \citep{faruna2026lsr} & HA & Native-authored (no native review) & Paired probes & 14 probes/lang; refusal $\sim$90\% EN vs.\ $\sim$40\% HA \\
    \midrule
    \multicolumn{5}{l}{\textit{SWAHILI (mid)}} \\
    AfriHate \citep{muhammad2025afrihate} & SW & Native-authored & Labelled posts & 21,092 tweets; $\kappa=0.55$ \\
    AfriSenti \citep{muhammad2023afrisenti} & SW & Native-authored & Labelled tweets & 2,401 (SemEval-2023 split, \citealp{muhammad2023semeval}; resource paper's Table 6 reports 3,014) \\
    PolitiKweli \citep{amol2024politikweli} & SW & Native-authored & Labelled tweets & 6,345 code-switched + 211 SW \\
    Political hate speech \citep{onyango2024swahili} & SW & Native re-annotation & Labelled tweets & 101,014 merged; $\kappa$ $0.435\rightarrow0.552$ \\
    XTREMESPEECH \citep{maronikolakis2022listening} & SW & Native-authored & Labelled passages & Kenya slice; $\kappa=0.13$; $\sim$8\% pure SW \\
    RTP-LX \citep{dewynter2024rtplx} & SW & Transcreated & Prompt-only & $\sim$1,100 prompts (V0.3, Jan 2024) \\
    MultiJail \citep{deng2024multilingual} & SW & Human-translated & Prompt-only & 315 prompts; $\sim$83\% unsafe (2023) \\
    UbuntuGuard \citep{abdullahi2026ubuntuguard} & SW & Synthetic + MT & Multi-turn & 1,899 train / 435 test; policy 93.30, transcript 96.99 \\
    Marx and Dunaiski (2026) re-test & SW & MT (derivative) & Single + multi-turn & Filtered subset; single translator \\
    Kiswahili agentic deep-dive \citep{intlnetwork2025agentic} & SW & Native + MT & Agentic trajectories & 156 tasks per language (boundary case, \S\ref{sec:scope}) \\
    \midrule
    \multicolumn{5}{l}{\textit{FRENCH (high)}} \\
    Aya Red-teaming \citep{aakanksha2024multilingual} & FR & Native-authored & Prompt-only & $\sim$900 prompts; global/local labels \\
    Jigsaw Multilingual \citep{jigsaw2020multilingual} & FR & Native (organic) & Labelled comments & 10,920 test comments; eval-only \\
    RTP-LX \citep{dewynter2024rtplx} & FR & Transcreated & Prompt-only & $\sim$1,100 prompts (V0.1, Sept 2023) \\
    XSafety \citep{wang2024all} & FR & Translated + proofread & Prompt-only & $\sim$2,800 entries; 14 categories \\
    M-ALERT \citep{friedrich2024llms} & FR & Translated & Prompt-only & $\sim$15,000 prompts; 6 macro / 32 micro \\
    CultureGuard \citep{joshi2025cultureguard} & FR & Synthetic + adapted + MT & Prompt+response & $\sim$43k samples; Aegis 2.0 taxonomy \\
    PolyGuardMix \citep{kumar2025polyguard} & FR & Mostly MT & Prompt+response & 1.91M total / 17 languages; per-language counts not published \\
    PolygloToxicityPrompts \citep{jain2024polyglotoxicityprompts} & FR & Native (web-scraped) & Prompt-only & 25k prompts; no human harm labels \\
    \bottomrule
  \end{tabular}
  \caption{The audited corpus by resource tier. Provenance and size are recorded for the language slice, not the parent collection. Where a paper and a live release disagree on size, both are shown. The Ubisoft/ToxBuster resource is reviewed and excluded (\S\ref{sec:boundarydecisions}) and does not appear here or in any count.}
  \label{tab:corpus}
\end{table*}

\subsection{Provenance composition}

Classifying each slice by how it was produced gives the distribution in Table~\ref{tab:provenance}. A slice is counted once, under its dominant production method as stated in its source paper. The Kiswahili agentic slice, which combines a minority of natively-authored Kenyan fraud tasks with a majority of machine-translated benchmark tasks, is counted under translation on that dominant-method basis, matching its Native + MT provenance in Table~\ref{tab:corpus}.

\begin{table}[t]
  \centering\small
  \begin{tabular}{lccccc}
    \toprule
    \textbf{Tier} & \textbf{Nat.} & \textbf{Trans.} & \textbf{Transl.} & \textbf{Synth.} & \textbf{Slices} \\
    \midrule
    Hausa (low) & 5 & 0 & 1 & 1 & 7 \\
    Swahili (mid) & 5 & 1 & 3 & 1 & 10 \\
    French (high) & 3 & 1 & 3 & 1 & 8 \\
    \bottomrule
  \end{tabular}
  \caption{Provenance composition by tier. Counts are slices, not datasets. ``Trans.'' denotes transcreated (translation with native-speaker cultural adaptation), distinguished from plain human ``Transl.'' translation because the source papers themselves draw that distinction.}
  \label{tab:provenance}
\end{table}

Read as raw counts, the three tiers look similar: each has some native material and some translated material. This apparent similarity is the reason a provenance count alone is insufficient, and why \S\ref{sec:provenance} separates provenance by harm domain. The native material is not comparable across tiers. For Hausa, three of the five native slices (AfriHate, HOC, and NaijaOffens) carry hate or offensive-speech labels; a fourth, AfriSenti, is native Hausa tweet data but a sentiment corpus with no harm labels, so it counts toward provenance only; the fifth, LSR, natively authors adversarial probes, but the harm they encode is individually targeted physical violence, which our extremism category, defined around ideological or communal violence, does not admit (\S\ref{sec:harmtax}, \S\ref{sec:boundarydecisions}). For Swahili, four of the five native slices carry harm labels (AfriHate, PolitiKweli, the Onyango set, and XTREMESPEECH), covering hate, offensive speech, political misinformation, and incitement, while AfriSenti again supplies native provenance without harm labels. For French, native material includes a purpose-built red-teaming set spanning four of our six harm categories with explicit global-versus-local harm labelling.

\subsection{Format and domain composition}

By format, the corpus is dominated by single-turn prompt-only and labelled-item resources. Multi-turn dialogue data exists for Hausa and Swahili through a single synthetic pipeline, and for French through one prompt-and-response guard-training corpus. This matters because of the downstream finding reported in the literature we build on: single-turn translation-based jailbreaks have been substantially mitigated in recent models, while multi-turn conversational attacks remain effective. The data that would be needed to evaluate the remaining vulnerability is, in our corpus, almost entirely synthetic and machine-translated for the two African languages.

By domain, the corpus is heavily concentrated on Twitter and X. Every native Hausa and Swahili slice except one draws on it. The exceptions are informative: one Hausa dataset supplements Twitter with Facebook material collected with dictionary guidance, and the French corpus is the only one in our study drawing on materially different domains, including Wikipedia talk pages and web-crawl text; an industrial game-chat resource (Ubisoft's ToxBuster line) was reviewed but excluded because its data is not public. Domain monoculture is therefore itself stratified by resource level, with consequences for distribution shift that we return to in \S\ref{sec:discussion}.

By script, every slice in our corpus is Latin. No resource covers Ajami, the Arabic-derived script used for Hausa, despite the literature acknowledging its use. Code-switching, by contrast, is modelled explicitly in four slices (Hausa-English in one benchmark component, Swahili-English in three), which makes it one of the few dimensions where the African-language resources are ahead of the French ones.

\subsection{Corpus boundary decisions}
\label{sec:boundarydecisions}

\S\ref{sec:scope} set three exclusion criteria: general-purpose resources with no safety purpose, method or model papers that release no new data, and lexicons or wordlists carrying no labelled text. Applying those criteria, together with the flag-and-boundary rules in the same section, the corpus resolves into what we dropped and what we kept under qualification. Two resources were dropped outright. NaijaOffens \citep{aliyu2024beyond} satisfies every inclusion criterion but was never released: its repository advertises 17,150 annotated tweets behind a data directory that does not exist, so it can support no availability-dependent analysis. Ubisoft's ToxBuster line \citep{yang2023toxbuster,yang2025unified} was reviewed and dropped because its classifier is published but its underlying game-chat data is proprietary; it contributes no slice and enters no count. Three resources were kept but qualified. PolyGuard \citep{kumar2025polyguard}, the largest multilingual guard-training corpus at 1.91M samples, covers seventeen higher-resource languages that include French but neither Hausa nor Swahili; we retain it for French and record its absence for the African tiers as a finding, not an omission. AfriSenti \citep{muhammad2023afrisenti} is retained but flagged as repurposed under \S\ref{sec:scope}: it is native tweet data for both Hausa and Swahili and counts toward provenance, but it is a sentiment corpus with no harm labels and does not count toward harm coverage. The Kiswahili agentic-safety resource is retained as a boundary case and excluded from the harm-coverage matrix (\S\ref{sec:boundary}), because its fraud and information-leakage categories map into none of our six harm categories. One further slice, LSR \citep{faruna2026lsr}, is in scope as adversarial-safety data but sits partly outside the harm matrix: its Hausa probes are natively authored and count toward provenance, yet the harm they encode is individually targeted physical violence rather than the ideological, organised, or communal violence our extremism category is defined around (\S\ref{sec:harmtax}), so we do not count it toward native extremism coverage.

\section{Findings}
\label{sec:findings}

We organise findings by gap category, following \S\ref{sec:gapdef}, with resource tier as the cross-cutting comparison inside each. \S\ref{sec:quantquality} comes first among the categories because it is the one case in our corpus where a single pipeline permits a controlled comparison, and it therefore calibrates the interpretation of everything that follows. \S\ref{sec:tracking} assesses whether gap severity tracks resource level, which is the paper's central question.

\subsection{Overview: gap incidence by tier}

Table~\ref{tab:gapseverity} summarises the assignment described in \S\ref{sec:methodology}. The pattern is neither uniformly monotonic nor random. Two of the six categories, provenance and access and licensing, worsen steadily as resource level falls; taxonomy coverage leaves the low tier clearly worst but the high and mid tiers level; the within-pipeline quality comparison (\S\ref{sec:quantquality}) runs the same way but spans only two of the three tiers; annotation and agreement is worst at the mid tier; and reuse and double-counting concentrates by research community rather than by tier.

\begin{table*}[t]
  \centering\footnotesize
  \begin{tabular}{p{2.6cm}p{4.2cm}p{4.2cm}p{4.2cm}}
    \toprule
    \textbf{Gap category} & \textbf{Hausa (low)} & \textbf{Swahili (mid)} & \textbf{French (high)} \\
    \midrule
    Provenance gaps & Severe: all multi-harm data translated or synthetic & Moderate: native present but single-domain & Mild: native red-team data exists, but breadth is translated \\
    Annotation / agreement gaps & Mixed: strong where reported ($\kappa$ 0.66--0.75), absent in half & Severe: lowest agreement observed ($\kappa$ 0.13--0.55) & Moderate: slice-level human agreement printed for few French slices (PolyGuard $\alpha \approx 0.47$); most report none \\
    Access \& licensing gaps & Severe: unreleased, request-only, unlicensed, password-gated & Moderate: mostly public; one no-repo case & Mild: mostly open; one industrial set never released \\
    Taxonomy-coverage gaps & Severe: native coverage confined to hate and harassment & Moderate: adds misinformation and incitement natively & Mild: all six covered, mostly via inherited taxonomies \\
    Reuse / double-counting gaps & Mild: limited overlap (AfriSenti $\rightarrow$ AfriHate) & Severe: four-way re-annotation chain & Mild: shared English seeds across benchmarks \\
    Quantified quality gaps & Documented deficit: policy MT 66.37 (below threshold); transcripts 93.31 & Documented advantage: policy 93.30, transcripts 96.99 & Not measurable: the one within-pipeline comparison (\S\ref{sec:quantquality}) does not include French \\
    \bottomrule
  \end{tabular}
  \caption{Gap severity by category and tier. Severity levels are assigned per slice against the anchoring criteria in \S\ref{sec:methodology} and aggregated to the tier by the modal and worst-case slice.}
  \label{tab:gapseverity}
\end{table*}

\subsection{Quantified quality gaps: a controlled within-pipeline comparison}
\label{sec:quantquality}

The strongest evidence in our audit for a resource-level quality gap comes from a single 2026 resource that processed Hausa and Swahili through an identical pipeline: the same expert-written English seed queries, the same models generating policies and dialogues, the same machine-translation stage, the same native-validation protocol, and the same automatic translation-quality scoring with the same acceptance threshold. Because the pipeline is held constant, the difference between the two languages isolates translation and validation quality from confounds of dataset origin, taxonomy, annotator pool, or research intent.

The result is a large and precisely located gap. On the paper's own translation-quality metric, Swahili scores 93.30 on policy material and 96.99 on transcripts, and Hausa scores 93.31 on transcripts, all comfortably above the authors' stated acceptance threshold of 70. The Hausa policy material alone scores 66.37, below the threshold the authors themselves set. The deficit is therefore not a general Hausa translation failure: conversational transcripts translated well, and the structured policy documents did not. The below-threshold slice was nonetheless retained and released as part of a resource whose collection-level framing presents ten African languages as comparably covered.

Three features make this case unusually probative. First, the deficit is measured by the authors, not inferred by us; we are reporting a number the paper publishes about itself. Second, the threshold is the authors' own, so the judgment that this material is below standard is not an external imposition. Third, the validation effort was equal in nominal terms (a single native validator reviewing twenty sampled pairs for each language), which means the equal treatment produced unequal reliability, precisely the mechanism by which a resource can be procedurally fair and substantively unequal at the same time.

\paragraph{Why this matters beyond one dataset.} Recent work identifies translation quality, rather than language identity as such, as the dominant factor explaining residual cross-language variation in jailbreak success. If that mechanism holds, then a measured translation-quality deficit for the lowest-resource language in a benchmark is not a documentation blemish; it is a direct predictor that safety evaluation in that language will be less reliable than the benchmark's coverage claim implies. Our audit finds the deficit exactly where that literature predicts it will matter most.

\subsection{Provenance gaps}
\label{sec:provenance}

Provenance gaps are the most frequent category in our corpus and the one that most clearly stratifies by resource level, but the stratification appears only when provenance is crossed with harm domain rather than counted in aggregate.

\paragraph{Hausa.} Native provenance exists and is not weak, but it is narrow. Three of the five native Hausa slices (AfriHate, HOC, and NaijaOffens) carry hate or offensive-speech labels; a fourth, AfriSenti, is native Hausa tweet data but a sentiment corpus with no harm labels, so it supports provenance without adding harm coverage. The fifth, LSR, is the one native Hausa resource whose content lies beyond hate and offensive speech: its adversarial probes were written natively, though by a single author and with native-speaker review explicitly listed as absent. The harm those probes encode is individually targeted physical violence, which our extremism category does not admit (\S\ref{sec:harmtax}, \S\ref{sec:boundarydecisions}), so LSR adds no native harm-category coverage; its headline result is nonetheless gap evidence, with refusal near 90\% for English probes and roughly 40\% for Hausa. Every other Hausa resource covering non-hate harms is translated or synthetic, and two carry documented provenance overstatements: TukaBench's African-language prompts were written in English and machine-translated before native post-editing, contrary to descriptions implying native composition, and UbuntuGuard's expert-authorship claim applies only to English seed material, as its paper states.

\paragraph{Swahili.} Swahili is the only African-language tier in our study with native provenance extending beyond hate and offensive speech, reaching political misinformation and, partially, incitement. It is also the only tier represented on both sides of the provenance divide by genuinely independent resources: native corpora on one side, transcreated and human-translated benchmarks on the other. This makes Swahili the most informative case for the native-versus-translated comparison, and it is the empirical basis for our treatment of it as intermediate rather than low, notwithstanding the CommonCrawl-share convention discussed in \S\ref{sec:scope}.

\paragraph{French.} French shows the mildest provenance gap but not the absence of one, and the shape of the gap is the more interesting result. French has purpose-built native red-teaming data covering four of six harm categories with global-versus-local harm labelling, and organic native toxicity data from a non-social-media domain. Yet the breadth of French harm coverage rests on translated and synthetic benchmarks, and one French native resource, PolygloToxicityPrompts, carries no human harm annotation at all, relying on an automated toxicity API by its authors' own acknowledgement. A high-resource language whose safety-evaluation breadth still depends on translation is a distinct class of finding, and it is the reason we resist reading our results as a simple scarcity story.

\subsection{Annotation and agreement gaps}
\label{sec:agreement}

Agreement reporting in our corpus is sparse and, where present, frequently poor. We count a slice as reporting slice-level agreement only when its source paper prints a numeric inter-annotator agreement statistic for that specific language's own labels. Six of the 25 slices qualify: AfriHate for Hausa ($\kappa=0.75$) and for Swahili ($\kappa=0.55$), AfriSenti for Hausa ($\kappa=0.66$), XTREMESPEECH for Swahili ($\kappa=0.13$), the \citet{onyango2024swahili} political hate-speech set for Swahili ($\kappa$ rising from 0.435 to 0.552 across rounds), and PolyGuard for French (Krippendorff's $\alpha \approx 0.47$ on safety labels). The rule sets aside three things that are easy to miscount as agreement: a collection-level figure not broken out by language, such as RTP-LX's single overall weighted $\kappa$, whose per-language values appear only as unlabelled heatmaps and in its release repository rather than as printed numbers; model-judge-versus-human agreement, such as TukaBench's 69.4\% for Hausa, which measures a different thing; and a calibration exercise standing in for agreement, such as UbuntuGuard's single-validator threshold check. The remaining slices report no agreement of any kind.

The distribution of reported agreement does not follow resource level in the direction one might expect. Where Hausa agreement is reported it is comparatively strong: 0.75 in AfriHate and 0.66 in AfriSenti, both free-marginal measures in native corpora. The weakest agreement in the entire audit belongs to XTREMESPEECH's Kenya slice at $\kappa = 0.13$, in a resource that is otherwise among the most methodologically careful in our corpus: it recruits local fact-checkers precisely to address annotator-community mismatch, and it publishes a full data statement. The low figure is reported honestly and discussed by its authors.

\paragraph{Interpretation.} We read low agreement in this setting as evidence about taxonomies, not about annotators. When trained local annotators cannot agree on whether a passage is derogatory, exclusionary, or dangerous, the most economical explanation is that the category boundaries do not correspond to a distinction the annotator community shares. This reading is reinforced by the fact that agreement is highest in the corpora whose taxonomies were designed by and for the language community, and lowest where a graded severity scale developed for cross-country comparison is applied to a specific national context.

\paragraph{The ethics-driven absence.} In at least one case, agreement is unreported for a defensible reason: prolonged annotator exposure to abusive material is itself a harm, and limiting redundant annotation limits that exposure. We take this seriously and do not code it as negligence. It nonetheless has a structural consequence that the field should confront directly, since the languages for which annotator welfare constraints bite hardest are the languages with the smallest annotator pools, with the result that the least-resourced slices end up least verified, for reasons that are locally ethical and globally inequitable.

\subsection{Access and licensing gaps}

Access gaps concentrate sharply at the low-resource tier. For Hausa, NaijaOffens satisfies every quality criterion but was announced for release and never shipped; HOC is request-only with no stated license and no reported size; TukaBench's live release contains 300 items for Hausa where its paper describes 986, because the two AfriJail components have not been uploaded. For Swahili, the \citet{onyango2024swahili} political hate-speech resource had no immediately locatable repository until its Zenodo and Science Data Bank deposits were traced through the article itself. For French, access is largely open, with one exception, ToxBuster, whose training data is proprietary. That is a failure in the opposite direction: proprietary, not under-resourced.

\paragraph{The paper-versus-release gap as a distinct failure mode.} We separate this from licensing because it is invisible to the checks a downstream developer is likely to run. A citation resolves, a repository exists, a dataset card renders, and the artifact is nonetheless not what the paper describes. Verifying it requires opening the release and counting, which is exactly the step that a collection-level reading skips. In our corpus this failure mode occurs only for the low-resource language.

Documentation completeness follows the same gradient. Annotation guidelines are fully public for a minority of slices, partially available for most, and absent for several, with absence concentrated in the Hausa resources. HOC reports no size at all: we verified across three versions of the paper that the figure is genuinely never stated, which we record as a documentation finding, not an incomplete audit.

\subsection{Taxonomy-coverage gaps}
\label{sec:taxonomygaps}

Crossing our six harm categories with native availability produces the matrix in Table~\ref{tab:harmcoverage}, which is the most direct answer this paper offers to the question of what is actually covered.

\begin{table}[t]
  \centering\small
  \begin{tabular}{p{1.7cm}p{1.55cm}p{1.65cm}p{1.65cm}}
    \toprule
    \textbf{Harm category} & \textbf{Hausa (low)} & \textbf{Swahili (mid)} & \textbf{French (high)} \\
    \midrule
    Hate speech & Native, multi-dataset & Native, multi-dataset & Native + translated \\
    Harassment & Native (``abusive'', ``offensive'') & Native (``abusive'', derogatory) & Native + translated \\
    Self-harm & None & Transcreated only & Native (Aya) + translated \\
    Extremism & None & Native (partial: incitement) & Translated only \\
    Sexual content & Translated only & Transcreated only & Native (Aya) + translated \\
    Misinformation & Synthetic / translated only & Native (PolitiKweli) & Translated only \\
    \bottomrule
  \end{tabular}
  \caption{Harm coverage by category and tier, distinguishing native from translated and transcreated provenance. ``Native'' requires that items in the target language were written or organically collected in that language, not translated into it.}
  \label{tab:harmcoverage}
\end{table}

Two results stand out. First, native harm-category coverage is four categories for French, four for Swahili, and two for Hausa. The high and mid tiers are level rather than separated, so the clean gradient a resource-level account would predict is absent at the top; only the low tier is clearly behind. Second, self-harm and sexual content have no native coverage in either African language, and for Hausa self-harm there is no coverage of any kind: no native, translated, transcreated, or synthetic resource in our corpus carries a self-harm label for Hausa. Where these two categories reach Hausa or Swahili at all, they arrive only through translated or transcreated benchmarks, so the culturally specific ways in which these harms are expressed, euphemised, or signalled in those languages go unrepresented. Given that these are precisely the categories where idiom and indirection carry the signal, this is the coverage gap we would prioritise.

The ordering is also not monotonic. Misinformation has native coverage in Swahili, through an election-driven code-switched corpus, and no native coverage in French. A high-resource language therefore has a native coverage gap that a mid-resource language does not, because coverage follows local research and civic priorities, not resource level alone. This is the clearest counterexample in our audit to a pure scarcity account, and it constrains how strongly the resource-level thesis in \S\ref{sec:tracking} can be stated.

Taxonomy imposition compounds the coverage gap. The dominant taxonomies in our corpus (a ten-behaviour jailbreak scheme, a twelve-category hazard hierarchy, a six-macro-category safety scheme, and an automated toxicity API's operational definition) originate in English-language frameworks and are inherited wholesale by the translated resources. Genuinely local harm conceptualisation exists but is a minority practice: global-versus-local harm labelling in one native red-teaming set, hate-target attributes covering communal and political categories in an African hate-speech collection, a culturally-grounded prompt subset in one benchmark, and a national-context policy grounding in one synthetic pipeline. Notably, three of these four exceptions are African-language resources, which suggests the imposition problem is being addressed from within the affected communities, not by the collections that most loudly claim multilingual coverage.

\subsection{Reuse and double-counting gaps}

Apparent corpus diversity in our audit is materially lower than the slice count suggests, and the compression is concentrated in Swahili. AfriHate's Swahili slice is assembled by re-annotating three earlier corpora: the negative class of the AfriSenti sentiment set \citep{muhammad2023afrisenti}, the PolitiKweli election-misinformation corpus \citep{amol2024politikweli}, and the Hate\_Speech\_Kenya corpus of \citet{ombui2019hate}. Separately, the \citet{onyango2024swahili} political hate-speech resource merges PolitiKweli, AfriSenti, and the same Ombui et al.\ corpus into a single reported set of 101,014 items. The same underlying pool of tweets therefore spreads across four nominally distinct entries in our corpus, AfriSenti, PolitiKweli, AfriHate, and the Onyango set, each carrying a different taxonomy, so a reader counting datasets counts the same text more than once.

A second, subtler form of reuse operates through shared English seeds. Several French benchmarks derive from a common pool of English red-teaming prompts and toxicity seeds, so that what presents as independent corroboration across benchmarks is partly the same source material translated by different pipelines. Reuse in the French case inflates apparent methodological independence; reuse in the Swahili case inflates apparent data volume.

\paragraph{Consequences.} Three follow. First, any count of items available for Swahili that sums across our corpus overstates the underlying pool substantially. Second, a model evaluated on two of these resources has not been independently evaluated twice. Third, and more constructively, the re-annotation chain is a natural experiment: the same items labelled under three taxonomies reveal how much of a harm judgment is carried by the taxonomy rather than the text. We note this as an opportunity in \S\ref{sec:recommendations} as well as a hazard.

\subsection{The boundary case: safety coverage outside the framework}
\label{sec:boundary}

One Swahili slice in our corpus, the Kiswahili agentic-safety material contributed by Kenya AISI to the International Network's third joint testing exercise, is native-reviewed, methodologically serious, produced by a national safety institute, and addresses agentic harms, fraud and sensitive-information leakage, through multi-turn tool-use trajectories. None of its harm categories map into our six-category framework. Under \S\ref{sec:harmtax} it is recorded as out-of-framework and excluded from Table~\ref{tab:harmcoverage}.

We retain it because it demonstrates a failure mode that a coverage count cannot express. A downstream reader who observes that a national institute has conducted Kiswahili safety evaluation may reasonably conclude that Kiswahili safety has been assessed. That conclusion would be wrong in a specific way: the content-safety and jailbreak gap this paper documents is left entirely untouched by that work, and the two facts are compatible. The resource also reports Kiswahili as one of three languages, with Telugu and Hindi, for which the model-judge assessments diverged most from human ones, reproducing in a different harm domain the same verification pattern we find in ours.

\subsection{Does gap severity track resource level?}
\label{sec:tracking}

Our central question admits a qualified answer: gap severity tracks resource level cleanly for two of the six gap categories, leaves the low tier clearly worst but the high and mid tiers level for a third, is directional but measurable across only two tiers for a fourth, is worst at the mid tier for a fifth, and is distributed by mechanism rather than tier for the sixth. We state the qualifications because the unqualified claim is not what our data show.

\paragraph{Where the gradient holds.} Provenance and access and licensing worsen monotonically from French through Swahili to Hausa: native provenance narrows and access failures multiply at each step down the ladder, and access failures serious enough to prevent use occur only for Hausa. Taxonomy coverage points the same way at the bottom but not the top. The number of our six harm categories with any native coverage is four for French, four for Swahili, and two for Hausa, so the low tier is clearly behind while the high and mid tiers are level. The one controlled within-pipeline comparison available to us (\S\ref{sec:quantquality}) runs in the same direction, placing Hausa's policy material below its own paper's quality threshold and the Swahili slice from that same pipeline well above it, though it compares only two of our three languages.

\paragraph{Where it inverts.} Agreement quality is worst at the mid tier, not the low tier, and native misinformation coverage exists for the mid-resource language and not the high-resource one. Both inversions have the same explanation: coverage and annotation practice follow the priorities and capacity of the research communities working on a language, not the language's web-crawl share. Swahili's low agreement figure comes from a resource attempting something genuinely hard, graded severity judgments across national contexts, while its misinformation coverage comes from a locally motivated response to a specific election.

\paragraph{Where the gradient is the wrong frame.} Reuse and double-counting is not a resource-level phenomenon but a research-community one: it concentrates wherever a small number of foundational corpora are repeatedly re-annotated, which happens to be Swahili in our corpus and could equally happen at any tier.

The finding we would emphasise is therefore not that low-resource languages have less safety data, which is unsurprising, but that the deficits compound in a specific and largely invisible way. For Hausa, the same slices that are narrowest in harm coverage are also the ones most likely to be unreleased, unlicensed, undocumented as to size, and, where translated, measurably below their own pipeline's quality bar. A reader assessing coverage from collection-level claims sees none of this, because each individual claim is technically accurate. It is the conjunction that fails, and only slice-level auditing surfaces the conjunction.

\section{Discussion}
\label{sec:discussion}

\subsection{The central claim}

\S\ref{sec:tracking} shows that gap severity tracks resource level cleanly for only two of six categories, leaves the low tier worst but the high and mid tiers level for taxonomy coverage, inverts for one, and is better explained by research-community structure than by resource level for another. This is a more precise and more defensible claim than ``low-resource languages have less safety data,'' and it is worth being explicit about why the more precise claim is the one this paper is actually making. A pure scarcity account predicts a monotonic gradient across every dimension we measured. It does not predict that Swahili's inter-annotator agreement would be the worst in the corpus, or that French would have zero native misinformation coverage while Swahili has native coverage driven by a specific election. Both of these results come directly out of \S\ref{sec:agreement} and \S\ref{sec:taxonomygaps}. A resource-level account that cannot accommodate them is not wrong so much as insufficiently mechanistic: it describes an outcome without describing the process that produces it.

The process our findings point to is that safety data provision is shaped by what a given research community has been able to build, which is a function of resource level but not reducible to it; it is also a function of what harms that community has had occasion to prioritize (an election, in the case of Swahili misinformation), what ethical constraints its annotation practice operates under (\S\ref{sec:agreement}'s annotator-welfare finding), and which foundational corpora happen to exist for it to build on top of (\S\ref{sec:findings}'s reuse chains). Resource level sets the ceiling on how much independent, well-verified data a community can produce; it does not determine what that community chooses to produce within that ceiling, or how visible the gaps in it will be to an outside reader.

\subsection{The conjunction problem}

The finding we would ask a reader to carry away from this paper, if only one, is the one stated at the close of \S\ref{sec:tracking}: for Hausa, the same slices that are narrowest in harm coverage are also the ones most likely to be unreleased, unlicensed, undocumented as to size, and, where translated, measurably below their own pipeline's quality bar; each of these facts, read in isolation from a paper's abstract or a dataset card, is individually unremarkable. A paper stating ``this dataset covers hate speech in Hausa'' is not lying. A paper stating ``translation quality exceeded our threshold for eight of nine languages'' is not lying either, even when Hausa is the ninth. It is the conjunction of narrow coverage, thin verification, restricted access, and marginal quality co-occurring in the same slice that constitutes the actual risk to a downstream reader, and no single field in a dataset card is designed to expose a conjunction. This is, we think, the most transferable methodological point in the paper: auditing dataset quality one claim at a time will systematically understate risk in exactly the cases where risk is highest, because the failure mode is not any one claim being false but the correlation structure across claims being invisible from where a consumer of the dataset is standing.

\subsection{From dataset gaps to model behavior}

\S\ref{sec:intro} and \S\ref{sec:related} establish, via \citet{deng2024multilingual} and \citet{marx2026multilingual}, that single-turn translation-based jailbreak attacks in low-resource African languages have been substantially mitigated between 2023 and 2026, while multi-turn conversational attacks in the same languages remain effective at rates comparable to English. Our findings offer a structural explanation for exactly this asymmetry, rather than merely an analogous or coincidental one. \S\ref{sec:corpus} shows that multi-turn dialogue data exists for Hausa and Swahili through a single synthetic, machine-translated pipeline, and that no native-authored multi-turn resource exists for either language anywhere in our corpus. Single-turn prompt data, by contrast, is the format in which the majority of native Hausa and Swahili material we found was produced. If the safety training and evaluation data available to a model provider for a given language and format shapes how well that provider can harden the model against that attack surface (which is the working assumption behind every dataset in our corpus) then the asymmetry in what has and has not been patched is not a coincidence running parallel to our findings; it is the outcome our findings would predict. The attack surface with abundant, comparatively well-verified native data (single-turn hate/offensive-content detection) is the one that has closed. The attack surface with only synthetic, machine-translated, single-validator multi-turn data (\S\ref{sec:quantquality}'s UbuntuGuard case is the only in-scope multi-turn safety resource we found for either language, and it is itself synthetic and machine-translated; the two other multi-turn items in our corpus, the Marx and Dunaiski re-test and the Kiswahili agentic slice, are a derivative re-test and an out-of-framework boundary case) is the one that remains open.

We state this as a structural correspondence, not a demonstrated causal chain: we do not have access to what data any specific commercial provider used to train or evaluate their models, and the correspondence could in principle run in the other direction, with providers under-investing in multi-turn African-language safety data because the attack surface was not yet publicly demonstrated to be exploitable, rather than the attack surface remaining exploitable because the data was thin. Distinguishing these directions is beyond what a dataset audit alone can establish. What the audit does establish is that if a provider wanted to close this specific gap today, the raw material to do so with any confidence (native-authored, multi-turn, harm-diverse, agreement-verified African-language safety data) does not yet exist in the published literature, which is itself an actionable finding regardless of how the historical causality resolves.

\subsection{Two counterexamples that constrain the paper's claim}

We flag two results from \S\ref{sec:findings} that a reader should hold in mind as bounds on how far the resource-level thesis generalizes, because ignoring them would make the paper's central claim easier to state but less true.

First, self-harm and sexual content have zero native coverage in either Hausa or Swahili, and Hausa self-harm has no coverage of any kind (\S\ref{sec:taxonomygaps}, Table~\ref{tab:harmcoverage}), a gap that is total rather than gradated and that does not distinguish between the low and mid tier at all. This is the single most concerning result in the paper on safety-relevance grounds, since these are precisely the categories in which local idiom, euphemism, and indirection are likely to carry the signal that a translated prompt set would miss. It is also the result least explained by a resource-level gradient, since French coverage in these categories is itself thin and translation-dependent (\S\ref{sec:provenance}). This suggests the gap here is less about resource level and more about a general under-prioritization of these two harm categories across the field's safety-data effort, African-language or otherwise.

Second, PolygloToxicityPrompts, a French, high-resource-tier, natively-collected resource, carries no human harm annotation at all (\S\ref{sec:provenance}, \S\ref{sec:corpus}). A resource-level account of dataset quality predicts that high-resource languages accumulate the most rigorous, human-verified safety data; this case shows that scale and native provenance do not guarantee human verification even at the high-resource tier, and that ``high-resource'' should not be read as a proxy for ``fully verified'' any more than ``low-resource'' should be read as a proxy for ``absent.''

\subsection{Relation to prior calls for better documentation}

\citet{bender2018data} and \citet{gebru2021datasheets} argue that a dataset's fitness for use cannot be assessed from a headline description and propose structured documentation as the remedy. Our findings are consistent with that diagnosis and extend it in one specific direction: even where the documentation these frameworks call for is present (several datasets in our corpus include a full data statement, and XTREMESPEECH is a strong example of exactly the practice Bender and Friedman recommend), the documentation is typically produced and consumed at the collection level. A data statement covering four countries and four languages answers Bender and Friedman's questions once, for the collection as a whole, and a reader inferring per-language reliability from it is doing exactly the extrapolation our audit shows to be unsafe (\S\ref{sec:findings}'s XTREMESPEECH Kenya slice, at $\kappa = 0.13$, sits inside a resource whose collection-level documentation is otherwise exemplary). Good documentation practice, as currently practiced, is a necessary but not sufficient condition for the kind of verification this paper performs; \S\ref{sec:recommendations} proposes what closing that remaining gap would require in practice.

\section{Recommendations}
\label{sec:recommendations}

We organize recommendations by audience, since the action available to a dataset creator differs from what is available to a model provider citing someone else's dataset, and both differ from what a venue or funder can require. Each recommendation traces to a specific gap category from \S\ref{sec:gapdef}/\S\ref{sec:methodology} so that its rationale is visible rather than asserted.

\subsection{For dataset creators}

\noindent\textbf{R1. Report at the slice level by default.} Per-language size, provenance, and format should be reported as a matter of course, not reconstructed by an auditor from scattered tables, as was necessary for several resources in our corpus (\S\ref{sec:verification}, \S\ref{sec:corpus}). Where a collection spans multiple languages, a single per-language table (comparable to the one we compiled as Table~\ref{tab:corpus}) should be standard, not exceptional.

\noindent\textbf{R2. Distinguish provenance precisely}, and audit your own claims against your own methodology section before publication. \S\ref{sec:verification} identified two cases in our corpus where a resource's secondary description implied native authorship that its own methodology section contradicted. This is avoidable: authors are in the best position to check that their abstract's provenance claim matches their methods section's actual account, and doing so before submission would have caught both cases we found by external audit.

\noindent\textbf{R3. Report inter-annotator agreement at the slice level, or state explicitly why not.} Where agreement is withheld for annotator-welfare reasons (\S\ref{sec:agreement}), we do not think this should be discouraged, but the omission itself, and the specific reason for it, should be stated in the paper rather than left for a reader to notice its absence. A one-sentence statement (``slice-level IAA was not computed for [language] to limit annotator exposure to abusive material; validation instead used [alternative method]'') converts a silent gap into a disclosed, defensible design choice, and is consistent with how HOC already handles this in our corpus; we would like to see this become the norm rather than the exception.

\noindent\textbf{R4. Verify that the released artifact matches the paper's description} before the paper is finalized, and update either the paper or the release if they diverge after publication. \S\ref{sec:agreement}'s access section paper-versus-release gap (TukaBench's 300 released items against 986 described; NaijaOffens never shipped at all) is the failure mode most invisible to a downstream reader, because checking for it requires opening the repository and counting rather than reading the paper. We recommend this become a checklist item at submission and, where a release is later updated, a note in the paper's repository documenting what changed and when.

\noindent\textbf{R5. State the license and access terms explicitly}, even if the answer is ``unclear'' or ``request-only.'' Several resources in our corpus provide no stated license at all, which is a distinct and worse condition than a restrictive license, since it leaves a downstream user unable to determine whether use is permitted rather than merely constrained.

\noindent\textbf{R6. When merging or re-annotating existing corpora, disclose the full provenance chain.} Our reuse findings were only traceable because the merging resources cited their sources; we ask that this become more prominent (for instance, a table showing exactly which prior corpus contributed which subset) rather than a citation buried in prose, both to prevent inadvertent double-counting downstream and to make the re-annotation itself more legible as a research contribution in its own right.

\subsection{For model providers and benchmark consumers}

\noindent\textbf{R7. Do not cite collection-level coverage statistics as evidence of per-language safety without slice-level verification.} A model card stating evaluation ``across 14 languages'' should specify, for any language a claim is being made about specifically, what that language's actual slice looked like (size, provenance, and verification status) rather than inheriting the collection's aggregate framing. Table~\ref{tab:gapseverity} is offered as a template for what this kind of disclosure could look like in practice.

\noindent\textbf{R8. Treat multi-turn, native-authored safety data as a distinct and currently underserved requirement}, not a natural extension of single-turn coverage. \S\ref{sec:discussion}'s structural correspondence between data format and residual jailbreak vulnerability suggests that closing the single-turn gap does not transfer to the multi-turn attack surface, and provider safety roadmaps should treat the two as requiring separately sourced data rather than assuming one substitutes for the other.

\noindent\textbf{R9. Weight self-harm and sexual-content coverage as a priority gap independent of overall resource level}, given \S\ref{sec:taxonomygaps} and \S\ref{sec:discussion}'s finding that this gap is total rather than gradated and is not well explained by the low/mid/high framing that organizes the rest of this paper.

\subsection{For venues, reviewers, and funders}

\noindent\textbf{R10. Consider requiring a per-language reporting table as a submission checklist item} for any dataset paper claiming multilingual coverage, in the spirit of existing reproducibility and data-statement checklists, but specifically targeting the slice-versus-collection distinction this paper argues is currently unenforced.

\noindent\textbf{R11. Fund native-authored, multi-turn, agreement-verified data collection specifically}, rather than machine-translation-and-post-edit pipelines, for the languages and harm categories this audit finds least covered: self-harm and sexual content in Hausa and Swahili, and multi-turn conversational data in both. \S\ref{sec:quantquality}'s controlled comparison shows that even well-resourced, methodologically careful synthetic pipelines can fall below their own quality threshold for a lower-resource language on structurally complex material (policy documents, in that case); this is evidence that synthetic-and-translate pipelines are not a substitute for native collection at the frontier of coverage, however useful they are for filling volume in the interim.

\section*{Limitations}
\label{sec:limitations}

\paragraph{Single-annotator audit.} As noted in \S\ref{sec:methodology}, gap-severity assignments were made by a single annotator per slice, with flagged rows reviewed jointly by the author team but without independent double-coding of the full corpus. We report the anchoring criteria precisely so that our assignments can be contested or reproduced, but we have not measured our own inter-rater reliability, which is a direct methodological tension given that unmeasured annotation reliability is one of the gap categories we critique in the datasets we review. We note this tension explicitly rather than let a reader discover it unaddressed.

\paragraph{Three languages, one tier boundary in dispute.} Our low/mid/high framing rests on three languages, and \S\ref{sec:scope} already discloses that Swahili's tier assignment is contested within our own corpus depending on whether resource level is operationalized by CommonCrawl share or by the practical availability of curated safety resources. Findings that depend on Swahili's position specifically (most notably \S\ref{sec:agreement}'s agreement inversion and \S\ref{sec:taxonomygaps}'s misinformation counterexample) should be read as evidence about the specific research-community dynamics around Swahili safety data, and not automatically generalized to every language a CommonCrawl-based classification would place in the same tier.

\paragraph{Corpus discovery is not exhaustive}, and is itself subject to a version of the paper's own critique. Our corpus was assembled through citation-following and targeted search rather than a systematic, pre-registered literature review, and is therefore susceptible to the same visibility bias the paper argues affects downstream readers of individual datasets: resources published in venues, or in languages, less central to the English-language NLP citation graph may be underrepresented in our own corpus for reasons structurally similar to the ones we identify in \S\ref{sec:findings}. We consider it likely, rather than merely possible, that safety-relevant Hausa- or Swahili-language resources exist that our search did not surface, and we would characterize our corpus as a substantial and structured sample rather than a census.

\paragraph{Severity categories are ordinal judgments, not measurements, for five of six gap types.} Only \S\ref{sec:quantquality}'s quantified-quality-gap category rests on a metric independent of our own judgment. The remaining five categories (provenance, agreement, access, taxonomy coverage, reuse) are assigned against the anchoring criteria in \S\ref{sec:methodology}, which we designed to be as observable and citable as possible, but which still require a judgment call at the boundary between, for instance, ``moderate'' and ``severe.'' We have tried to make every such judgment traceable to a specific paper section or repository state rather than an impression, but we do not claim the four-level scale itself is a measurement instrument with known validity.

\paragraph{Findings are time-bound.} Several resources in our corpus are explicitly incomplete as of the time of review, most notably TukaBench's undeployed AfriJail components (\S\ref{sec:corpus}, \S\ref{sec:findings}), and dataset releases, once updated, are not guaranteed to remain in the state we audited. We date our findings to the access dates recorded per slice in our instrument and would expect a re-audit conducted even a year later to find a different, likely improved, picture for at least some of the specific access and release gaps we report, even if the structural pattern we argue for persists.

\paragraph{We did not independently validate our own harm-taxonomy mapping against native-speaker judgment.} \S\ref{sec:methodology}'s mapping of native taxonomies onto our six-category framework is a documented judgment made by the author team from each source paper's stated definitions, not verified against an independent native-speaker reading of the underlying items. Given that \S\ref{sec:agreement} argues that low measured agreement in our source corpora may itself reflect taxonomy mismatch rather than annotator unreliability, we hold our own taxonomy-mapping judgments to the same standard of scrutiny, and would welcome exactly the kind of independent verification we are unable to provide within the scope of this study.

\section*{Ethics Statement}
\label{sec:ethics}

\paragraph{No new harmful content was created.} This paper does not collect, generate, or publish new instances of hate speech, offensive language, self-harm content, or other harmful material. Our contribution is an audit of already-published datasets' documentation, methodology, and release state; where we reproduce examples from source papers, these are drawn from material the original authors already published with their own content warnings, and we do not introduce new offensive examples beyond what is necessary to illustrate a specific, cited finding.

\paragraph{This paper is a systemic critique, not an assessment of individual researchers.} Several of the gaps we document, most visibly the annotator-welfare-driven absence of inter-annotator agreement discussed in \S\ref{sec:agreement}, arise from defensible, good-faith responses to real constraints (funding, annotator availability, ethical obligations to annotators) rather than from carelessness. Several of the resources we audit most closely (AfriHate, AfriSenti, XTREMESPEECH, HOC) are themselves unusually transparent about their own limitations, and our ability to identify their gaps at all is partly a function of that transparency, not evidence that these resources are worse than less transparent alternatives we may have under-scrutinized as a result. We have tried throughout to frame findings at the level of the field's incentive structure and resourcing pattern, and we ask readers, including reviewers, to resist reading this paper as a ranked judgment of specific research teams.

\paragraph{We are conscious of the risk that this work could be misused to justify disinvestment rather than reinvestment.} A paper documenting that low-resource-language safety datasets are thinner and less verified than their high-resource counterparts could, read uncharitably, be used to argue that such datasets are not worth building or citing. This is the opposite of our intended reading. Every gap we document is a gap in available data, not a demonstration that closing it is not worthwhile or not possible; \S\ref{sec:quantquality}'s controlled comparison, in particular, shows a below-threshold Hausa slice sitting inside a pipeline whose Swahili output cleared the same bar easily, which is evidence that the gap is addressable within existing methods given sufficient native-speaker validation effort, not evidence that Hausa safety data collection is intrinsically harder to do well.

\paragraph{No human subjects were recruited for this study.} All data reviewed was drawn from already-published papers and public or previously-accessed repositories; we did not conduct new annotation, translation, or user studies of our own.

\paragraph{Data and audit-instrument availability.} We release the completed audit instrument, all 25 language slices, scored against the thirty-one-field schema described in \S\ref{sec:methodology} and the six gap categories described in \S\ref{sec:gapdef}/\S\ref{sec:methodology}, together with the per-field rationale recorded for every judgment, as supplementary material accompanying this paper. Our intent is that a reader be able to verify any specific severity judgment reported in \S\ref{sec:findings} directly against the evidence and reasoning that produced it, rather than accept our summary tables on trust. This is the same standard of disclosure we ask dataset creators to meet in \S\ref{sec:recommendations} (R1: slice-level reporting by default; R5: explicit access and licensing terms), and we did not think it consistent to hold the datasets we audit to a transparency standard we were unwilling to apply to our own work. Where our own judgments are contestable, and \S\ref{sec:methodology}/\S\ref{sec:limitations} already acknowledge that four of our six gap categories rest on ordinal assessment rather than measurement, releasing the underlying instrument allows a reader to identify exactly where they disagree with us and why, which we regard as a more useful outcome than a table whose reasoning is not inspectable.

% Acknowledgments (safe to include now that this is the non-anonymous
% preprint version). Remove for an anonymous review submission.
\section*{Acknowledgments}
We thank Alisar Mustafa for her mentorship and guidance throughout this work. This work was conducted independently by the authors under Black in AI Safety \& Ethics (BASE).

\bibliography{refs}

\end{document}